\newif\ifelsevier
\elsevierfalse

\ifelsevier
\documentclass[11pt]{elsart}
\def\CAL{\mathcal}
\else
\documentstyle[11pt]{article}
\def\CAL{\cal}
\fi

\def\re#1{(\ref{#1})}
\def\vgl#1{eq.\ (\ref{#1})}

\newcommand\T{\theta_{12}}
\newcommand\Tb{{\bar\theta}_{12}}
\newcommand\Z{Z_{12}}
\newcommand\D{{\CAL D}}
\newcommand\Db{\overline{\CAL D}}
\def\proj{{\CAL P}}
\def\NO#1{:\!#1\!:}
\def\range#1{{{[}#1{]}}}

\def\ie{{\sl i.e.\ }}
\def\eg{{\sl e.g.\ }}
\newcommand{\bin}[2]{\left(\!\!\!\begin{array}{c}#1\\#2\end{array}\!\!\!\right)}

\newcounter{mathline}
\newcounter{mathref}

\def\un{\symbol{95}}
\def\ha{\symbol{94}}
\newlength{\tab}
\newcommand{\inm}[1]{\hspace*{11pt}\addtocounter{mathline}{1}
    \makebox[\tab][l]{\sl In[\themathline] :=}{\tt #1}}
\newcommand{\outm}[1]{\hspace*{11pt}
    \makebox[\tab][l]{\sl Out[\themathline] =}{\tt #1}}
\newcommand{\contm}[1]{\hspace*{11pt}
    \makebox[\tab][l]{}{\tt #1}}
\newcommand{\contmind}[1]{\hspace*{11pt}
    \makebox[\tab][l]{}{\hspace*{.8cm} \tt #1}}
\newcommand{\Mathematica} {{\sl Mathematica}}
\newcommand{\SOPEdefs} {{\sl SOPEN2defs}}
\newcommand{\OPEdefs} {{\sl OPEdefs}}
\newcommand{\Dummies} {{\sl Dummies}}

\begin{document}
\ifelsevier

\begin{frontmatter}

\title{A {\sl Mathematica\raise.6ex\hbox{\normalsize TM}} Package for
Computing $N=2$ Superfield Operator Product Expansions}

\author[Dubna,Frascati]{S. Krivonos\thanksref{emailSK}}
\author[IC]{K. Thielemans\thanksref{emailKT}}

\address[Dubna]{
JINR -- Bogoliubov Laboratory of Theoretical Physics,\\
141980 Dubna, Moscow Region, Russia.}
\address[Frascati]{
INFN-Laboratori Nazionali di Frascati,\\
 P.O. Box 13, I-00044 Frascati, Italy.}
\address[IC]{
Theoretical Physics Group, Imperial College, \\
London SW7 2BZ, UK.}

\thanks[emailSK]{Email address : krivonos@thsun1.jinr.dubna.su}
\thanks[emailKT]{Email address : k.thielemans@ic.ac.uk}

\else

\renewcommand{\thefootnote}{\fnsymbol{footnote}}
\topmargin -1cm
\textheight 8.5in
\textwidth 6in

\thispagestyle{empty}

\begin{flushright}
                   Preprint Imperial/TP/95-96/13 \\
                   hep-th/9512029
\end{flushright}

\vspace{8mm}

\begin{center}
{\large\bf A {\sl Mathematica\raise.6ex\hbox{\normalsize TM}} Package for
Computing $N=2$ Superfield Operator Product Expansions}
\footnote{
This work was partly carried out in the framework of the project ``Gauge
theories, applied supersymmetry and quantum gravity'', contract
SC1-CT92-0789 of the European Economic Community.
}
\\[1cm]

{\bf S. Krivonos}\footnote{Email address : krivonos@thsun1.jinr.dubna.su}
\\*[2mm]
JINR -- Bogoliubov Laboratory of Theoretical Physics,\\
141980 Dubna, Moscow Region, Russia.\\
{\em and}\\
INFN-Laboratori Nazionali di Frascati,\\
 P.O. Box 13, I-00044 Frascati, Italy.
\\[.5cm]
{\bf K.\ Thielemans}\footnote{Email address : k.thielemans@ic.ac.uk}
\\*[2mm]
Theoretical Physics Group\\
Imperial College, London SW7 2BZ, UK.
\\[.5cm]
November 1995
\end{center}
\vspace*{.2cm}
\fi

\begin{abstract}
We describe a general purpose \Mathematica${}^{TM}$ package for
computing  Superfield Operator Product Expansions in meromorphic $N=2$
superconformal field theory. Given the SOPEs for a set of ``basic"
superfields, SOPEs of arbitrarily complicated composites can be computed
automatically.  Normal ordered products are always reduced to a standard
form. It is possible to check the Jacobi identities, and to compute
Poisson brackets (``classical SOPEs''). We present two explicit examples:
a construction of  the ``small'' $N=4$ superconformal algebra in terms
of $N=2$ superfields, and a realisation of the $N=2$ superconformal
algebra in terms of chiral and antichiral fermionic superfields.
\end{abstract}

\ifelsevier
\end{frontmatter}

\else
\newpage
\renewcommand{\thefootnote}{\arabic{footnote}}
\setcounter{footnote}0
\setcounter{page}1

\fi

\section{Introduction}
It is well known that Operator Product Expansions (OPEs) form a powerful
tool in two--dimensional Conformal Field Theory (CFT). Once the OPEs
for the currents are known, it is possible to compute their
correlation functions algebraically. One also derives differential
equations for the other correlation functions in the theory \cite{BPZ}.
In gauge theories, one can compute the anomaly in the BRST nilpotency,
find the spectrum, etc. Although the calculations are in principal
straightforward. In practice, they can become quite cumbersome to do by hand.
In the case of a meromorphic CFT, these calculations can be performed
with the \OPEdefs\ package \cite{OPEdefs2.0,KTthesis},
implemented in \Mathematica${}^{TM}$ \footnote{ \Mathematica\ is a
trademark of Wolfram Research Inc.  For details, see \cite{Wolfram}.}, an
interactive environment for performing symbolic computations.

In the case of supersymmetric conformal theories, it is often useful
to work in a superspace formalism. This drastically reduces the number
of fields one needs to consider, and also the amount of algebraic
manipulations. For example, a general $N=2$ superfield contains 4
components fields (two bosonic and two fermionic), therefore one SOPE
for the $N=2$ superfield defines 10 OPEs for the components. Moreover,
the checking of one Jacobi identity in terms of $N=2$ superfields is
equivalent to 64 identities in terms of components.

Unfortunately, although the formulas to deal with super OPEs are
similar to the ones for ordinary OPEs, they are more complicated and
calculations are more error--prone. This paper describes a \Mathematica\
package that enables one to perform computations automatically in the
case of $N=2$ supersymmetry.

Of course, the simplest supersymmetric extension of the conformal field
theories possesses $N=1$ supersymmetry. But an $N=1$
superfield contains only one additional fermionic component with respect
to the bosonic case and so it is possible to use the \OPEdefs\ package. On
the other hand, when dealing with higher supersymmetry $N\geq 3$, most
of the superfields are constrained, but they can be nicely represented
in terms of $N=2$ ones. In the case of $N=2$, the constraints (chirality
or antichirality) can be easily taken into account in \SOPEdefs. This is
why we decided to limit ourselves for the time being to the case of
$N=2$ superfields.

The package is of a similar nature to \OPEdefs . One needs to declare the
superfields which are used, and give their respective SOPEs.
Nonlinear algebras can be handled by using a point--splitting
definition for composites. It is then possible to check the Jacobi
identities for the SOPEs. If one introduced arbitrary constants in the
SOPEs, it is possible in this way to find the values for which the
algebra is associative, and hence to construct a super $W$--algebra.
In this way, the superfield version of $N=2$ $W_3^{(2)}$ was constructed
in \cite{N2W32}.
Once the basic SOPEs are given, it is in principle possible to compute any
OPE (or a particular pole) of any composite. Of course, this is restricted by
the available memory and CPU--time.
The package is able to compute SOPEs, but also Poisson brackets
(``classical'' SOPEs). It has been tested by a number of people,
already resulting in many publications.

The package can be extended to perform calculations using dummy indices.
This is still a limited facility. Not all possible simplifications are
performed when tensors appear which have symmetries when interchanging
indices. However, for many cases this possibility is quite useful.

This paper is setup as follows. In a first section we explain our
conventions for $N=2$ superspace. Then we derive the formulas for SOPEs.
The next section consists of a user's guide for the package. We end with
two explicit examples.

\subsubsection*{Notation}
Input for and output from \Mathematica\ is written in {\tt typeset} font.
Input lines are preceded by ``{\sl In[n] :=}", and corresponding output
statements by ``{\sl Out[n] =}", as in \Mathematica.

\section{$N=2$ superspace. Notations and conventions}
In this section we will introduce our conventions concerning the $N=2$
superfield formalism.

The $N=2$ superspace $Z=\left\{ z,\theta,\bar\theta \right\}$ can be
described by one real bosonic coordinate $z$ and pair of two conjugate
Grassman cordinates $\theta,\bar\theta$:
\begin{equation}
 \theta^2 = \bar\theta^2 = 0, \qquad \theta\bar\theta = -\bar\theta\theta\,.
\end{equation}
The integration measure in
$N=2$ superspace is  $dZ\equiv dz d\bar\theta d\theta$ and the standard
convention for the integration over $\theta,\bar\theta$ is assumed.

To deal with $N=2$ superfields $\Phi (Z)$, it is useful to introduce chiral
and antichiral spinor derivatives $\D,\Db$
\begin{equation}
\D=\frac{\partial}{\partial \theta}-\frac{1}{2}\bar\theta
    \frac{\partial}{\partial z} ,\quad
\Db = \frac{\partial}{\partial \bar\theta} -\frac{1}{2}\theta
     \frac{\partial}{\partial z} , \label{DDB}
\end{equation}
which obey the following relations
\begin{equation}
{\D}^2=\Db{}^2=0, \quad
\left\{ \D,\Db \right\} = -\partial_z .
\end{equation}
Now we can define chiral $F$ and antichiral $\bar F$ superfields which
are subjected to the following constraints
\begin{equation}
\D F = \Db {\bar F} = 0,
\end{equation}
respectively.

It is also instructive to write down the Taylor expansion for a general
superfield $\Phi (Z)$
\begin{equation}
\Phi (Z_1) = \sum_{n \geq 0} \frac{Z_{12}^n}{n!}\partial_2^n\left\{
 1+\theta_{12}\D_2 +{\bar\theta}_{12}  \Db_2 -
 \theta_{12}\bar{\theta}_{12}\frac{1}{2}
 \left[ \D_2,\Db_2\right]\right\}\Phi (Z_2),
\label{eq:Taylor}
\end{equation}
where
\begin{equation}
\theta_{12}=\theta_1-\theta_2,\quad
{\bar\theta}_{12}={\bar\theta}_1-{\bar\theta}_2, \quad
Z_{12}=z_1-z_2+\frac{1}{2}\left( \theta_1{\bar\theta}_2-
      \theta_2{\bar\theta}_1 \right).
\end{equation}

We give the SOPE for the simplest case of $N=2$ superconformal algebra (SCA):
\begin{eqnarray}
J(Z_1)J(Z_2)&=&\frac{c/4}{Z_{12}^2} +
     \frac{\theta_{12}{\bar\theta}_{12}J(Z_2)}{Z_{12}^2}+\nonumber \\
&&  \frac{{\bar\theta}_{12}\Db J(Z_2)}{Z_{12}}-
    \frac{\theta_{12}{\D} J(Z_2)}{Z_{12}}+
    \frac{\theta_{12}{\bar\theta}_{12}\partial J(Z_2)}{Z_{12}},
\label{eq:N=2T}
\end{eqnarray}
where $J(Z)$ is a general bosonic superfield \cite{Schoutens}.

\section{Formulas for SOPEs}
In principle, the formulas to work with SOPEs follow by expanding
everything in components, applying the standard OPE formulas, and
reassembling again into superfields. However, this approach is
impractical. It is easier (and conceptually clearer) to construct
the formulas by using Taylor expansions, contour integration \ldots in
superspace. We will discuss SOPEs here and briefly comment on which changes
are needed for the Poisson bracket case.

In fact, it will be useful to introduce a notation valid for any number
of supersymmetries $N$ \cite{Schoutens}. For a list of numbers
$\range{i_l}$, we will write $\theta^\range{i}$ for
$\theta^{i_1}\theta^{i_2}\ldots$ and $\theta^\range{N}$ is used for
$\theta^{i_1}\ldots \theta^{i_N}$. The ``complement'' $N-\range{i}$ is
defined such that
$\theta^\range{N}=\theta^{N-\range{i}}\theta^\range{i}$.

We also introduce projectors $\proj$ to extract a component of
superfield or a SOPE. $\proj_\emptyset$ projects on the term without
$\theta$'s, $\proj_\range{1}$ recovers the term proportional to
$\theta^1$ and so on:
\begin{equation}
\proj_\range{i} = \int d\theta^\range{N}\ \theta^{N-\range{i}}
\label{eq:defproj}
\end{equation}

We will use the following notation for SOPEs:
\begin{equation}
A(Z_1) B(Z_2) = \sum_{n <= h(A,B)}
   {\frac{{[}AB{]}_n(Z_2,\theta_{12}^i)}{(\Z)^n}} \ \ ,
\label{eq:OPEdef}
\end{equation}
where $h(A,B)$ is some finite number, and is usually given in terms of
the conformal dimensions of $A$ and $B$ as $h(A,B) = [d_A + d_B + N/2]$.
Note that with this definition \re{eq:OPEdef}, $[AB]_n$ depends on the
$\theta_{12}^i$, but not on $z_1$.
The term in $[AB]_0$ which has no $\theta_{12}$ dependence is called the
normal ordered product of $A$ and $B$:
\begin{equation}
\NO{AB}\ =\ \proj_\emptyset[AB]_0\label{eq:defNO}\,.
\end{equation}

With the definition \re{eq:OPEdef}, we have:
\begin{equation}
\proj_\range{i}[AB]_n(Z_2,\T^i) =
\oint_{C_2} {dz_1\over 2\pi i} \int d\theta_1^\range{N}
   \ \theta_{12}^{N-\range{i}}(\Z)^{n-1}
 A(Z_1)B(Z_2)\,,\label{eq:OPEcontour}
\end{equation}
where the contour $C_2$ encircles $z_2$.
If we replace the SOPE of $A$ and $B$ by the Poisson bracket
$\{A(Z_1),B(Z_2\}_{PB}$, we adopt the above formula as the definition for
$[AB]_n$ with $n>0$ in the Poisson brackets (``classical'' SOPEs).

\subsubsection*{Formulas}
With all these definitions, we can start to derive how we have to
compute with SOPEs.
First, there are some formulas for computing an SOPE of two superfields when
one of them is a derivative. These follow simply by taking derivatives
of the SOPE. We give only one example (in $N=2$):
\begin{equation}
[\D\! A\, B]_n(Z_2,\theta_{12}^i) =
             {(n-1)\over 2} \Tb [A\, B]_{n-1}(Z_2,\theta_{12}^i) +
             \D_1 [A\, B]_n(Z_2,\theta_{12}^i)\,,
\end{equation}
where the subindex on $\D_1$ indicates it acts on coordinates $Z_1$.
This implies
\begin{equation}
\proj_\theta [A\, B]_{n} = \proj_\emptyset[ \D A\, B]_n\,.
\end{equation}
For a term in the regular part of an OPE, we find:
\begin{equation}
[A\, B]_{-n} = {1\over n!} [\partial^n A\, B]_0\ .\label{eq:regularPole}
\end{equation}

The SOPE $B(Z_1) A(Z_2)$ is by analytic continuation equal to $A(Z_2)
B(Z_1)$. It remains then to apply a Taylor expansion \re{eq:Taylor} to the
fields in the latter. The result is straightforward. We list only the
special case for normal ordering, where drastical simplifications occurs
because of the projection $\proj_\emptyset$ in \re{eq:defNO}:
\begin{equation}
\NO{BA}\ =\ (-1)^{|A||B|} \NO{AB} + (-1)^{|A||B|}
\sum_{l\geq 1}{\frac{(-1)^l}{l! } \partial^{l}\proj_\emptyset [AB]_l}
\label{eq:NOcomm}
\end{equation}
Applying this formula for $A=B$ a fermionic superfield, one gets an
expression for $\NO{AA}$ in terms of the poles in the SOPE $A(Z_1)A(Z_2)$.

Associativity of the SOPE means that SOPEs can be computed in any order
(inside correlation functions).
This can be used to derive the Jacobi identities. As a first step, we
select a particular term in a double SOPE by using contour integrals:
\begin{eqnarray}
\lefteqn{
\proj_\range{i}[A\, \proj_\range{j}[BC]_p]_q (Z_3) \, =
 \oint_{C_3} {dZ_1\over 2\pi i}\, \theta_{13}^{N-\range{i}}
 Z_{13}^{q-1}
 \nonumber}\\
&&\hspace*{3em}
        \oint_{C_3} {dZ_2\over 2\pi i}\,
\theta_{23}^{N-\range{j}}Z_{23}^{p-1}\,
A(Z_1)B(Z_2)C(Z_3)\,,
\label{eq:OPEcontoursR}
\end{eqnarray}
where $C_2$ denotes a contour which encircles $z_2$ once anti-clockwise.
We can now use a contour deformation argument relating the contour integral
in \vgl{eq:OPEcontoursR} to a contour integral where the integration over
$Z_2$ is performed last. In fact, the contour deformation works only for
the $\theta$--independent part of the integral. For the Berezin integral
over $\theta$, we just interchange the order of integration (taking
signs into account). The resulting integral has two
terms: one where the $z_1$ contour is around $z_3$, and one where it is
around $z_2$. We find:
\begin{eqnarray}
\lefteqn{\proj_\range{i}[A\, \proj_\range{j}[BC]_p]_q  \, =
  (-1)^{(|A|+|i|)(B+|j|)} \proj_\range{j}[B\,\proj_\range{i}[AC]_q]_p +
\nonumber}\\
&&\hspace*{3em} (-1)^{(|A|+|i|)(|j|)}
 \sum_{l>0}\bin{q-1}{l-1}
 \proj_\range{j}[\proj_\range{i}[AB]_l\, C]_{p+q-l}
    \label{eq:OPEJacRAB}\,.
\end{eqnarray}
This equation has to hold (inside correlators) for consistency of the
OPE-formalism. It is valid for {\sl any} integers $p,q$, \ie also
negative numbers. However, in practice we use the Jacobi identities
\vgl{eq:OPEJacRAB} for positive $p,q$ as equations for the singular
part of the SOPEs, while for $p$ or $q$ zero, they define how one should
calculate with composites. For example, to compute an SOPE with a
composite, we insert $p=0, \range{j}=\emptyset$ into
\vgl{eq:OPEJacRAB}. When $q=0, \range{i}=\emptyset$ as well, we find:
\begin{eqnarray}
\NO{A\ \NO{BC}\,} = (-1)^{|A||B|} \NO{B\ \NO{AC}\,} +
 \NO{\left(\NO{AB} -(-1)^{|A||B|}\NO{BA}\right)\ C} \label{eq:NOcompR} \ ,
\end{eqnarray}
where we used \vgl{eq:NOcomm}.

For Poisson brackets, the normal ordered product of two operators is
simply replaced by (noncommutative) multiplication. In particular, this
means that double contractions ($l<q$) in \vgl{eq:OPEJacRAB} for $p=0,
\range{j}=\emptyset$ drop out.
However, the Jacobi identities \vgl{eq:OPEJacRAB} for positive
$p,q$ remain exactly the same.

\subsubsection*{Improvements}
The rules given in this section up to now are sufficient to
compute any SOPE, and to reorder any composite into a standard form.
For example, to compute an SOPE where the first operator is a composite,
we could use the fact that $A(Z_1) B(Z_2)=B(Z_2)A(Z_1)$
and \re{eq:OPEJacRAB}. However, it is possible to construct a rule
by using contour integrals to do this in one step. We find:
 \begin{eqnarray} [\proj_\emptyset [AB]_0\ C]_q &=&
    \sum_{l\geq 0}{1 \over l!}  [\partial^l A\ [BC]_{l+q}]_0
   + (-1)^{|A||B|}\sum_{l\geq 0}{1 \over l!} [\partial^l B\ [AC]_{l+q}]_0
\nonumber\\
   & & + (-1)^{|A||B|} \sum_{l=1}^{q-1} [B\ [AC]_{q-l}]_l\,,
\label{eq:OPEcompL}
\end{eqnarray}
where $q \geq 1$ and we used \vgl{eq:regularPole}.
To have the formula in this simple form, we used the convention that any
terms with $\T,\Tb$ move to the left of the OPE bracket. For instance,
if $[BC]_m = \T D$, then $[A\ [BC]_m]_n] = (-1)^{|A|} \T [AD]_n$.

For composites, we find:
\begin{eqnarray}
\NO{\,\NO{AB}\  C} &=&
    \NO{A\ \NO{BC}\,} +\nonumber\\
&&\!\!\!\!\!\!\!\!\!
  \sum_{l>0} {1 \over l!} \NO{\partial^l A\ \proj_\emptyset[BC]_{l}}
    + (-1)^{|A||B|} \sum_{l>0}{1\over l!}
                   \NO{\partial^l B\ \proj_\emptyset[AC]_{l}} \,.
 \label{eq:NOcompL}
\end{eqnarray}

\section{User's Guide}
\setcounter{mathline}{0}
This section is intended as a user's guide to the package \SOPEdefs.
Explicit examples are given for most operations.
\SOPEdefs\ is based on \OPEdefs\ 3.1 \cite{KTthesis} and most commands
are exactly the same. \SOPEdefs\ requires \Mathematica\ 2.0 or later.

As \SOPEdefs\ is implemented as a \Mathematica\ package, it has to be
loaded {\em before} any of its global symbols are used. Loading the
package a second time will clear all previous definitions of operators
and OPEs, as well as all stored intermediate results. Assuming that the
package is located in the \Mathematica-path, \eg\ in your current
directory\footnote{Use the \Mathematica\ command {\tt SetDirectory} for
this.}, issue:
\\[2mm]
\inm{<<SOPEN2defs.m}
\\[2mm]
After loading \SOPEdefs\ into \Mathematica, help for all the global
symbols is provided using the standard help-mechanism, \eg {\tt ?OPE}.

Now, you need to declare the operators that will be used. If you
want to define bosonic (resp. fermionic) operators, use {\tt Bosonic}
(resp. {\tt Fermionic}). For chiral operators, prefix with {\tt Chiral},
for antichiral ones, use {\tt AChiral}. For instance, to define
a bosonic field {\tt J} and an antichiral bosonic field $W$,
the statements are:
\\[2mm] \inm{Bosonic[J]}
\\ \inm{AChiralBosonic[W]}
\\[2mm]
For derivatives we use the following notation:
\begin{equation}
\partial_z J \rightarrow \mbox{\tt J'}\,,\quad
\D J \rightarrow \mbox{\tt DT[J]}\,,\quad
\Db J \rightarrow \mbox{\tt DBT[J]}\,,\quad
[\D,\Db] J \rightarrow \mbox{\tt DDB[J]}
\end{equation}

The order of the declarations fixes also the
ordering of operators used by the program:
\begin{equation}
\mbox{{\tt J' < DBT[J] < DDB[J] < DT[J] < J < W
}} \label{eq:OPEdefsOrder}
\end{equation}
By default, spatial derivatives of an operator are
considered ``smaller'' than the operator itself. This can be reversed
using the global option {\tt NOOrdering} (see  below).

Finally, the nonregular OPEs between the basic operators have to be
given  by listing the operators  that occur at the poles, the first
operator in the list is the one at the highest non-zero pole, the last
operator has to be the one at the first order pole. For example, to give
{\tt J} the OPE \re{eq:N=2T},  and {\tt W}  of a conformal dimension $1$
primary antichiral superfield with  $U(1)$--charge $-2$:
\\[2mm]
\\   \inm{OPE[J,J] = MakeOPE[\{c/4 One + t12**tb12**J,}
\\ \contmind{ tb12**DBT[J]-t12**DT[J]+t12**tb12**J'\}];}
\\   \inm{OPE[J,W] = MakeOPE[\{t12**tb12**W,}
\\ \contmind{ -2 W - t12**DT[W]+t12**tb12**W'\}];}
\\[2mm]
Note the operator {\tt One} which specifies the unit-operator. The
symbols {\tt t12}, {\tt tb12} denote $\T,\Tb$. In an OPE, they should
always be multiplied with an operator using {\tt **}. In particular, one
could have {\tt t12**One}.
\begin{quote}
\noindent{\bf Warning}: it is important that the operators occuring as
arguments of {\tt OPE} in a {\em definition} should be given in the order the
operators are declared (\ref{eq:OPEdefsOrder}), otherwise wrong results will be
generated.
\end{quote}

A normal ordered product $\NO{A B}$ (\ref{eq:defNO}) is entered in the
form {\tt NO[A,B]}. Multiple composites can be entered using only one {\tt
NO} head, \eg {\tt NO[A,B,C]}. This input is effectively translated
into {\tt NO[A, NO[B, C]]}. All output is normal ordered with the same
convention, \ie from right to left (input can be in any order). Also,
the operators in composites will always be
ordered according to the standard order (\ref{eq:OPEdefsOrder}).
\\[2mm] \inm{OPE[J,NO[J,J]]}
\\ \outm{<< 4|| c t12**tb12**One/ 4 ||3|| 0 ||2|| c J/2 +}
\\ \contm{ tb12**DBT[J] + t12**DT[J] + 2 t12**tb12**NO[J, J]}
\\ \contm{ ||1|| tb12**DBT[J'] + 2 tb12**NO[J, DBT[J]] +}
\\ \contm{ t12**DT[J'] - 2 t12**NO[J, DT[J]] + }
\\ \contm{ 2 t12**tb12**NO[J', J] >>}
\setcounter{mathref}{\value{mathline}}
\\[2mm]
This also shows what the output for an OPE looks like. For an OPE
with $n$ poles, the output is {\tt << n|| ... ||n-1|| ... ... ||1|| ... >>}.

\begin{quote}
{\bf Warning}: when computing OPEs with composites, or when
reordering composites, \SOPEdefs\ remembers by default some
intermediate results. Thus, it is dangerous to change the definition of
the basic OPEs after some calculations have been performed. For example,
consider a constant $a$ in an OPE. If calculations are performed after
assigning a value to $a$, the intermediate results are stored with this
value. Changing $a$ afterwards will give wrong results.
\end{quote}
The other globally defined functions available from the package
are:
\begin{itemize}
\item \verb&OPEOperator[operator_, parity_]& provides a more general way to
declare an operator than {\tt Bosonic} and {\tt Fermionic}. The second
argument is the parity of the operator such that $(-1)^{\tt parity}$ is
$+1$ for a boson, and $-1$ for a fermion. It can be a symbolic constant.
This can be used to declare a $bc$-system of unspecified parity.
In such cases, the operator can contain a named pattern:
\\ \inm{OPEOperator[J[i\un],parity[i]]}\\
If one wants to declare more operators, one can group each operator and its
parity in a list:
\\ \inm{OPEOperator[\{b[i\un],parity[i]\},\{c[i\un],parity[i]\}]}\\
See also {\tt SetOPEOptions[ParityMethod, \un]}.
\item \verb&ChiralQ[operator_]& tests if \verb&operator& is chiral.\\
One can use \verb&ChiralQ[A] = True&.
\item \verb&AChiralQ[operator_]& is the same for antichiral operators.
\item \verb&OPEPole[n_][ope_]& gets a single pole of an OPE:
\\ \inm{OPEPole[2][Out[\themathref]]}
\\ \outm{c J/2 + tb12**DBT[J] + t12**DT[J] +}
\\ \contm{2 t12**tb12**NO[J, J]}\\
\verb&OPEPole[n_][A_,B_]& can also be used to compute only one pole term of
an OPE:
\\ \inm{\% - OPEPole[2][J, J]//Expand}
\\ \outm{0}\\
{\tt OPEPole} can also give terms in the regular part of the OPE:
\\ \inm{OPEPole[-1][J, J]}
\\ \outm{NO[J',J] + tb12**NO[DBT[J'],J] +}
\\ \contm{t12**NO[DT[J'],J] - t12**tb12**NO[DDB[J'],J]/2}
\\
\item \verb&MaxPole[ope_]& gives the order of the highest pole in the OPE.
\item \verb&OPEParity[A_]& returns an even (odd) integer if $A$ is
bosonic (fermionic).
\item \verb&OPESimplify[ope_, Function -> function_]& ``collects'' all
terms in {\tt ope} with the same operator and applies {\tt function} on the
coefficients. The default setting for the option {\tt Function} is {\tt
Expand},
but this can be changed using {\tt SetOptions}.\\
\verb&OPESimplify[pole_, Function -> function_]& does the same simplifications
on sums of operators.\\
The alternative syntax \verb&OPESimplify[ope_, function_]& is also allowed.
\item \verb&OPEMap[function_, ope_]& maps {\tt function} to all poles of
{\tt ope}.
\item \verb&OPEMapAt[function_, ope_, position_]& maps {\tt function} to
a part of {\tt ope} specified by {\tt position} (see {\tt MapAt}). For
example,\\
\contm{OPEMapAt[Expand, ope,\{1\}]}\\
maps {\tt Expand} on the highest order pole of {\tt ope}.
\item \verb&GetCoefficients[expr_]& returns a list of all
coefficients of operators in {\tt expr} which can be OPEs
or poles. Applied on a list, it maps on the elements of the list.
\item \verb&GetOperators[expr_]& returns a list of all
operators in {\tt expr} which can be OPEs
or poles. Applied on a list, it maps on the elements of the list.
\item \verb&OPEJacobi[op1_,op2_,op3_]& computes the
Jaco\-bi-iden\-ti\-ties \re{eq:OPEJacRAB} for the singular part of
the OPEs of the three arguments. In general, all different orderings
should be tried to ensure associativity.\\
{\tt OPEJacobi} returns a list of which all
should be zero up to null fields to be associative.\\
{\tt OPEJacobi} accepts an option {\tt Function -> function} which
it passes to {\tt OPESimplify} at intermediate steps.
\item \verb&Delta[i_,j_]& is the Kronecker delta symbol
$\delta_{ij}$.
\item \verb&Epsilon[i_,j__]& is the antisymmetric symbol (with any
number of arguments.
\item \verb&N2OPEToComponents[J_]& gives the $N=0$ components of a
superfield $J$. For $J= A + \theta B + \bar\theta C +
\theta \bar\theta D$ the components are $\{A,B,C,D\}$.
In \SOPEdefs, the components are represented as
\[ \{J,\D J,\Db J, -{1\over 2} [\D,\Db]J\}\,,\]
where a projection on the $\theta$ independent part is understood.\\
\verb&N2OPEToComponents[ope_,J1_,J2_]& computes the 16 OPEs of the
components of $J_1$ and $J_2$. It is a double list where the ($m,n$)-th
element is the $N=0$ OPE of the $m$-th component of $J_1$ with the
$n$-th component of $J_2$.
\item \verb&ClearOPESavedValues[]& clears all stored
intermediate results, but not the definition of the operators and
their OPEs. To clear everything, reload the package.
\item \verb&TeXForm[ope_]& gives \TeX\ output for
an OPE. The arguments are always $Z_1,Z_2$.
\item \verb&TeXFormTD& can be assigned a string which will be used by
{tt TeXForm} for the output of {\tt DT, DBT, DDB}.
\item \verb&OPESave[filename_]& (with {\tt filename} a string between
double quotes) saves the intermediate results that \OPEdefs\ remembers
to file (see the option {\tt OPESaving} below).
\item {\tt SetOPEOptions} is a function to set the global options of
the package. The current options are:
\begin{itemize}
\item \verb&SetOPEOptions[NOOrdering, n_]& :
if {\tt n} is negative, order higher spatial derivatives to the left
(default), if {\tt n} is positive, order them to the right.
\item \verb&SetOPEOptions[ParityMethod, 0|1]& : makes it possible to use
operators of an unspecified parity. When the second argument is $0$
(default), all operators have to be declared to be bosonic or
fermionic. When the argument is $1$, {\tt OPEOperator} can be used with
a symbolic parity.
Note that in this case, powers of $-1$ are used to compute signs, which is
slightly slower than the boolean function which is used by the first
method.\\
This option is not normally needed as the use of {\tt
OPEOperator} with a non-integer second argument sets this option
automatically.
\item \verb&SetOPEOptions[OPESaving, boolean_]& :
if {\tt boolean} evaluates to {\tt True} (default), \OPEdefs\
stores the intermediate results when computing OPEs of composites and
when reordering composites. This option is useful if \Mathematica\ runs
short of memory in a large calculation, or when computing with dummy
indices (see section \ref{ssct:example2}).
\item \verb&SetOPEOptions[OPEMethod, method_]& :
with {\tt method} set to {\tt Quan\-tum\-OPEs} enables
normal OPE computations (default setting), while {\tt Classi\-cal\-OPEs}
enables Poisson bracket computations. Using this option implicitly calls
the function {\tt Clear\-OPE\-Saved\-Values[]}.
\item \verb&SetOPEOptions[EnableDummies, True]& : loads the {\tt
Dummies`} package and sets it up.
\end{itemize}
\end{itemize}

\section{Examples}
\subsection{The $N=4$ superconformal algebra}
\setcounter{mathline}{0}
As a first example of using the {\sl SOPEdefs} package, let us construct the
OPEs of $N=4$ $SU(2)$ SCA \cite{Ademollo} in terms of $N=2$ superfields.
The $N=4$  $SU(2)$ SCA contains the following component currents: the
stress-tensor $T(z)$, affine $su(2)$ currents $J^i(z), i=1,2,3$ and two
doublets of fermionic spin 3/2 currents $G_{\alpha},{\bar{G}}{}^\beta,
\alpha,\beta=1,2$
which form respectively the fundamental and conjugated representations
of $su(2)$. Due to the spin structure, the simplest way to put these
component currents into $N=2$ supermultiplets is to start with a general
spin 1 $N=2$ superfield $J(Z)$ which forms the $N=2$ SCA (spin contents
is $\left\{ 1,3/2,3/2,2 \right\}$ together with two bosonic spin 1
chiral-antichiral superfields  $\bar W(Z),W(Z)$, ${\D}\bar W
={\Db}W=0$ (spin content  $\left\{ 1,3/2,1,3/2 \right\}$).

First of all we need to define all superfields we are dealing with:
\\ \inm{<<SOPEN2defs.m}
\\ \inm{Bosonic[J]}
\\ \inm{AChiralBosonic[W]}
\\ \inm{ChiralBosonic[WB] }\\
As the next step we define the OPE between  all the super currents:
\\   \inm{OPE[J,J] = MakeOPE[\{c/4 One + t12**tb12**J,}
\\ \contmind{ tb12**DBT[J]-t12**DT[J]+t12**tb12**J'\}];}
\\   \inm{OPE[J,W] = MakeOPE[\{t12**tb12**W,}
\\ \contmind{ -2 W - t12**DT[W]+t12**tb12**W'\}];}
\\ \inm{OPE[J,WB] =  MakeOPE[\{t12**tb12**WB,}
\\ \contmind{ 2 WB+tb12**DBT[WB]+t12**tb12**WB'\}];}
\\ \inm{OPE[W,WB] = MakeOPE[\{a1 t12**tb12**One,}
\\ \contmind{ a2 One + t12**tb12**J, a3 J + a4 t12**DT[J]\}];}\\
Let us note that due to the first OPE,  $J(Z)$ obeys the $N=2$ SCA,
while the second and third OPEs define $W(Z)$ and $\bar W(Z)$ as
antichiral-chiral superfields primary spin 1 with respect to $J(Z)$. As
concerning the last OPE we introduce the arbitrary coefficients $a1-a4$
for all the terms because we want to determine the exact form of this
OPE. The possible terms and their structure in this OPE are completely
fixed by the antichiral-chiral structure of the $W,\bar W$ supercurrents
(the coefficient before term $\T**\Tb**J$ can be chosen to be 1 due to
the scaling invariance $W\rightarrow \alpha W$).

Now we will fix all coefficients from the Jacobi identities
\footnote{
This statement takes 5 seconds on a Pentium 90 MHz running Mathematica 2.2
for Windows.}:\\
\inm{OPEJacobi[J,W,WB, Function -> Factor]//Union} \\
\outm{\{0, ((2-a3)*J)/4, ((2-a3)*J)/2, ((-2+a3)*J)/2,}\\
\contm{ (a3-a4)*J, ((a1-a2)*One)/2, (a1-a2)*One,}\\
\contm{ (-a1+a2)*One, ((4*a1+c)*One)/4, }\\
\contm{ ((8*a2+a3*c)*One)/4, ((8*a2+a4*c)*One)/4,}\\
\contm{ ((-2+a4)*DT[J])/2, (-a3+a4)*DT[J]\}}
\\
We can solve the system of equation inside \Mathematica, \eg
\\
\inm {sol1 = Solve[GetCoefficients[\%]==0]}\\
\outm{\{\{a1 -> -c/4, a2 -> -c/4, a3 -> 2, a2 -> 2\}\}}
\\
Now we can check that all other Jacobi identities are satisfied:
\\
\inm{OPESimplify[OPEJacobi[W,W,WB]/.First[sol1]]//Union }\\
\outm{ \{0\} }
\\
By this we finished the construction of the $N=4$ $SU(2)$ SCA in terms of
$N=2$ superfields.

\subsection{A Miura--like realisation of $N=2$}\label{ssct:example2}
\setcounter{mathline}{0}
As a second example we will demonstrate how to use the
\SOPEdefs\ package to find a realisation of the $N=2$ SCA in terms of
$M$ pairs of chiral--antichiral fermionic superfields (Miura type
realisation). We also show here our technique to deal with
summing over repeated indices, which is implemented in a separate
package \Dummies. This package can also be used together with \OPEdefs,
or even independently. We only explain some basic commands of \Dummies\ here.

The first step is loading the package.
\\ \inm{<<SOPEN2defs.m}\\
Enable working with dummy indices.
\\ \inm{SetOPEOptions[EnableDummies, True]}\\
We now define which indices we are going to use. All summation indices
will have the form {\tt i[1]}, {\tt i[2]}, etc. That is, they all have
{\tt Head} {\tt i}.
\\ \inm{DefineDummy[i]}\\
For convenience, we define an abbreviation for the summation range.
\\ \inm{dimension[i] = M;}\\
Now we define the superfields we are going to play with. In the
case at hand we can define $M$ pairs of chiral-antichiral fermionic
superfields $F_i,{\bar F}_j$ through two \Mathematica\ statements:
\\ \inm{ChiralFermionic[F[\un]]}
\\ \inm{AChiralFermionic[FB[\un]]}\\
Using indices the SOPEs between superfields $F_i$ and
${\bar F}^j$ can be defined mostly as they are defined in the
textbooks:
\\ \inm{OPE[F[i\un],FB[j\un]] := }
\\ \contmind{Delta[i,j] MakeOPE[\{-1/2 t12**tb12**One, One\}]}\\
Now we will introduce a composite superfield $J(Z)$ which should satisfy
the $N=2$ SCA \vgl{eq:N=2T}:
\begin{equation}
J = a_1 \sum_i \NO{ F_i{\bar F}^i} + a_2 \Db F_1
          + a_3 {\D}{\bar F}^1  \label{miura1}
\end{equation}
where besides the standard term $F_i{\bar F}^i$ we also introduced the
$N=2$ analog of a Feigin-Fuchs term. In \Mathematica\ this becomes:
\\ \inm{J := NewDummies[}
\\ \contmind{a1 NO[F[i[1]],FB[i[1]]] +}
\\ \contmind{a2 DBT[F[1]] + a3 DT[FB[1]]]}
\\
Note the {\tt NewDummies} statement and the assignment using `{\tt :=}'.
Together, they make sure that everytime {\tt J} is used, the summation
indices will have a new number.

After these definitions we are ready to do real calculations. For
example, to check for which values of parameters $a_1,a_2,a_3$ the
superfield $J$ \re{miura1} spans $N=2$ SCA, let us calculate the
SOPE $J$ with $J$ and subtract the known result \re{eq:N=2T}:
\\ \inm{OPESimplify[DummySimplify[}
\\ \contmind{OPE[J,J] - }
\\ \contmind{MakeOPE[\{c/4 One + t12**tb12**J,}
\\ \contmind{	 tb12**DBT[J]-t12**DT[J]+t12**tb12**J'\}]}
\\ \contmind{], Function -> Factor]}
\\ \outm{<< 2|| One (-c + 4 M a1\ha 2  - 8 a2 a3)/4 +}
\\ \contm{  (-1+a1) a2 t12 ** tb12 ** DBT[F[1]] +  ... ||1||}
\\ \contm{  (-1+a1) a1 tb12 ** NO[DBT[F[i[1]]], FB[i[1]]] +}
\\ \contm{  ...>>}
\setcounter{mathref}{\value{mathline}}
\\[2mm]
where {\tt DummySimplify} renumbers the dummy indices to get a unique
form\footnote{In fact, the result of {\tt DummySimplify} is not unique
when tensors with certain symmetries appear, \eg $J^{ij}=J^{ji}$.}
For brevity, we omitted a number of terms (all proportional to $a_1-1$)
in the output and replaced them by an ellipsis.
As can be immediately seen, the coefficient $a_1$ must
be equal to 1, while $a_2,a_3$ are arbitrary. Let us note that our result
{\em Out[\themathref]} also gives the expression for the central charge
of the $N=2$ SCA:
\begin{equation}
c= 4M-8a_2a_3
\end{equation}
Thus we reproduced the standard Miura realisation for the $N=2$ SCA.

We hope that these simple examples will help users to use
the \SOPEdefs\ package in more complicated calculations.

\section{How to get it, and the future}
If you are interested in \SOPEdefs, you can get it
by Email from the authors. You'll also find it at
\verb&http://euclid.tp.ph.ic.ac.uk/~krthie/&
or via anonymous ftp at {\tt euclid.tp.ph.ic.ac.uk}.
Please put a reference to this paper in your
paper when you use it. Questions, remarks
and improvements are welcome. The package wil be extended to other
numbers of supersymmetries $N$.

\ifelsevier
\begin{ack}
\else
\subsection*{Acknowledgements}
\fi
K.T. thanks the JINR for the pleasant welcome during the weeks when
this work was started. He also acknowledges the Instituut voor
Theoretische Fysica from the KU Leuven (Belgium) where the first
version of the package was completed. S.K. wishes to thank Imperial
College for the hospitality in the final stage of this work.
We also thank Lucy Wenham for improving our English.
\ifelsevier
\\
This work was partly carried out in the framework of the project ``Gauge
theories, applied supersymmetry and quantum gravity'', contract
SC1-CT92-0789 of the European Economic Community.
\end{ack}
\fi

\end{document}